\documentclass[%
 reprint,
 amsmath,amssymb,
 aps,
]{revtex4-1}

\usepackage{graphicx}
\usepackage{dcolumn}
\usepackage{bm}

\begin{document}

\preprint{APS/123-QED}

\title{Periodic Anderson Model with Holstein Phonons for the Description of the Cerium Volume Collapse}
\author{Enzhi Li$^{1,2}$,  Shuxiang Yang$^{1,2}$, Peng Zhang$^3$,
Ka-Ming Tam$^{1,2}$, Mark Jarrell$^{1,2}$, and Juana Moreno$^{1,2}$}
\affiliation{$^1$ Department of Physics \& Astronomy, Louisiana State University,
Baton Rouge, LA 70803, USA}
\affiliation{$^2$ Center for Computation \& Technology, Louisiana State University, Baton Rouge, LA 70803, USA}
\affiliation{$^3$ Department of Physics, Xi'An Jiaotong University, Xi'An, Shaanxi, China}


\date{\today}

\begin{abstract}
Recent experiments have suggested that the electron-phonon coupling may play an important role in the $\gamma \rightarrow \alpha$ volume collapse transition in Cerium. A minimal model for the description of such transition is the periodic Anderson model. In order to better understand the effect of the electron-phonon interaction on the volume collapse transition, we study the periodic Anderson model with coupling between Holstein phonons and electrons in the conduction band. We find that the electron-phonon coupling enhances the volume collapse, which is consistent with experiments in Cerium. While we start with the Kondo Volume Collapse scenario in mind, our results capture some interesting features of the Mott scenario, such as a gap in the conduction electron spectra which grows with the effective electron-phonon coupling.
\end{abstract}

\maketitle

\section{Introduction}
The isostructural volume collapse of Cerium is a long-standing puzzle~\cite{Cerium_1949}. When a crystal of Cerium is under a pressure of 15,000 atmospheres, it undergoes a volume collapse of approximately 17\% while preserving the face-centered cubic crystal structure. This transformation, called the $\gamma \rightarrow \alpha$ transition, has baffled physicists since its discovery, and several leading theories have been proposed for its explanation, the most prominent of which are the Mott transition scenario \cite{Mott_1974} and the Kondo volume collapse (KVC) scenario \cite{KVC_1982}. The Mott and KVC scenarios are competing paradigms, although perhaps not as different and distinct as previously thought~\cite{held2000similarities, held2001cerium}.

In the KVC scenario, the $4f$ electrons of Cerium are assumed to be localized in both phases. In the small volume $\alpha$ phase, the $spd$ electrons strongly screen the local moments of the $f$ electrons, thus rendering the $\alpha$ phase a Pauli paramagnet. While in the large volume $\gamma$ phase, the local moments of the $f$ electrons persist to much lower temperatures than in the $\alpha$ phase, indicating that the Kondo scale $T_{K}$ in the $\gamma$ phase is much smaller than that of the $\alpha$ phase, which is consistent with the experimental observations~\cite{PhysRevB.45.8934, PhysRevB.67.075108, PhysRevB.48.13981,PhysRev.149.551}. 

In the Mott transition scenario, for which the Hubbard model is a good description, the density of states (DOS) of the $f$ electrons changes from being metallic (no gap at the Fermi level)  in the $\alpha$ phase
to insulating (with a gap at the Fermi level) in the $\gamma$ phase~\cite{PhysRevLett.69.168, PhysRevB.45.6479, PhysRevB.47.3553, held2001cerium}. 
This localization-delocalization of the $4f$ electrons, which is a metal-insulator Mott transition, is driven by the increase of the intersite hopping amplitudes of the $f$ electrons when the unit cell volume of Cerium decreases. 

While there are extensive studies on the cerium volume collapse, there is no consensus on the mechanism of this transition. An overview on the cerium volume collapse can be found in Ref. \cite{Ce_review}. Most of the previous models
proposed consider exclusively the interplay among the $spd$ electrons and the $f$ electrons, whereas the possible effects from the phonons are completely ignored. A series of recent experimental results have indicated that the electron-phonon interaction may also play an important role in the $\gamma \rightarrow \alpha$ transition~\cite{phonon_2009, phonon_2011, entropy_2006,Jeong_2004}.  Jeong \textit{et al.}~\cite{Jeong_2004} estimated 
that about half of the entropy change during the transition is due to lattice vibrations.  Later, Krisch \textit{et al.}~\cite{phonon_2011}
showed that the significant changes in the phonon dispersion across the $\gamma\rightarrow\alpha$ transition provide strong evidences for the importance of the lattice degrees of freedom. Although the precise value of the lattice vibrational entropy varies between experiments, they do agree that a significant fraction of the total entropy change during the transition is due to lattice vibrations.
This calls for a revision of the previous models to incorporate the contribution from the electron-phonon coupling. 

Even if we focus exclusively on the electronic contribution, the full model should be more complicated than the simple single band Periodic Anderson or Hubbard models. 
Recent studies using density functional theory have confirmed that the $f-spd$ hybridization is important \cite{Casadei_2016,PhysRevLett.109.146402,Haule_2005,Chakrabarti_2014,PhysRevLett.111.196801}, while the smaller 
spin-orbit coupling and hybridization among the $f$ orbitals are two other contributions which should be taken into account for a quantitative description~\cite{Streltsov2010,Amadon_2015}. 

Parameters extracted from ab-initio band structure calculations have also been used as inputs for many-body methods which reveal  strong coupling effects which are presumably absent from the DFT, such as Kondo screening and Mott transition. These methods include dynamical mean field theory (DMFT) \cite{held2001cerium,PhysRevLett.111.196801, amadon2012self, Amadon_2015, PhysRevB.94.115148,Amadon_2015,Tian_2015}, variational Monte Carlo \cite{PhysRevB.91.081101}, and Gutzwiller projection approaches \cite{Tian_2015,Dong_2014}.
Constrained Random Phase Approximation have also been applied to estimate the coupling terms \cite{PhysRevB.89.125110,Amadon_2015}. 

Unfortunately, there is no well developed method to incorporate the electron-phonon interaction into these frameworks. Moreover there is no appropriate formalism to include dynamical phonons with non-trivial dispersions within the DMFT, as it involves effective non-local electron-electron interaction. A quantitative study of electron-phonon coupling with the accuracy on par with that of a computation with only electron-electron interactions is the ultimate goal but not practically feasible at present. In light of the difficulties on the modeling of electron-phonon coupling, we sought a model which is 
simple enough to handle computationally but nevertheless capture the first order transition. 
Electron-phonon coupling is then incorporated into the model and its effects are studied in some detail. 

To attain the above goal we consider phonons within the 
Kondo volume collapse scenario. This is in line with the original approach by Allen and Martin~\cite{KVC_1982}, who studied the electronic and the bulk modules contributions to the free energy. In their study, the electronic part is modeled
by a single impurity model. In this work, we use the periodic Anderson model with Holstein phonons coupled to the conduction band as our starting point~\cite{Peng_2013}.
We solve this model using the DMFT approximation with the continuous time quantum Monte Carlo as our impurity solver. We then use the maximum entropy method to extract the density of states (DOS) of the conduction electrons, and study the evolution of the DOS with varying parameters. 

Our main finding is that the electron-phonon interaction can significantly enhance the volume collapse, and associated with this collapse, a metal-insulator transition emerges. Although we start our model with the Kondo scenario in mind, yet a Mott metal insulator transition that manifests itself by the formation of a gap is observed. Thus, our work may pave the way for the unification of the competing Kondo and Mott scenarios, a unification that is already lurking in some previous works~\cite{held2000similarities, held2001cerium}. 

The structure of the paper is as follows. In section II, we briefly describe our model Hamiltonian and the methods we use to solve it. In section III, we present our results for the pressure-volume curves, the behavior of the DOS across the phase transition, and discuss the relationship between the volume collapse and the Mott transition. We presnt our conclusion in section IV. 

\section{Model and Method}
In order to study the influence of the phonons on the Cerium volume collapse, we have here employed the periodic Anderson model with electron-phonon interaction, which is  
\begin{eqnarray}
\hat{H} &=& \hat{H}_{0} + \hat{H}_{I} 
\label{eq:Hamiltonian}\\\nonumber
\hat{H}_{0} &=& -t\sum_{\langle i, j \rangle, \sigma} (c_{i,\sigma}^{\dagger} c_{j,\sigma} + c_{j,\sigma}^{\dagger} c_{i,\sigma}) + \epsilon_{f} \sum_{i, \sigma} f_{i, \sigma}^{\dagger}f_{i, \sigma} \\\nonumber
&& + V \sum_{i,\sigma} (c_{i,\sigma}^{\dagger}f_{i,\sigma} + f_{i,\sigma}^{\dagger}c_{i,\sigma}) + \sum_{i} \Big(\frac{P_{i}^2}{2m} + \frac{1}{2}k X_{i}^{2} \Big) \\\nonumber
\hat{H}_{I} &=& U\sum_{i} n_{i,\uparrow}^{f}n_{i,\downarrow}^{f} + g\sum_{i,\sigma} n_{i,\sigma}^{c} X_{i}\,.
\end{eqnarray}
where $c_{i, \sigma}^{\dagger}, c_{i, \sigma} (f_{i, \sigma}^{\dagger}, f_{i, \sigma})$ creates and destroys a $c(f)$ electron of spin $\sigma$ at lattice site $i$, respectively. $P_i$ and $X_{i}$ are the phonon momentum and displacement operators. Here, we have used dispersionless Einstein phonons with frequency $\Omega_0 = \sqrt{k/m}$. The parameter $g$ measures the electron-phonon interaction strength, $U$ is the Hubbard repulsion between localized $f$-electrons, and $V$ characterizes the hybridization between conduction- and $f$-electrons.
With the parameters $g$ and  $k$, we construct the effective electron-phonon interaction strength, $U_{eff} = \frac{g^2}{2k}$. 
Throughout this paper and to be consistent with the experimental results, we have set $\Omega_0 =0.01$~\cite{Peng_2013}
unless otherwise specified. 
To preserve the large temperature metallic phase, we fix the total electronic density at $n=1.8$ by tuning the chemical potential at each iteration of the DMFT cycle. We also choose an appropriate value for $\epsilon_f$  so that $n_f=1$ when $T=0.1$ to ensure that a local moment is present at high temperature. 
We set  $U = 4$, however, the precise value of $U$ is not crucial as we have found qualitatively similar results for other values of $U$. Since the strength of $U$ ( = 4) which measures the Hubbard repulsion strength between $f$ electrons is significantly larger than the values of $\epsilon_f$ ($\approx -0.1$) we use, most of the time, $f$ electron filling number is quite insensitive to the variations of $\epsilon_f$. For most of the parameters that we have scanned, we simply set $\epsilon_f = -0.15$ and assure ourselves that the $f$ filling will almost always be nearly 1.0 when $\beta = 10$. 
We use a hypercubic lattice with Gaussian bare DOS, and consider its bandwidth as our unit of energy.We set this unit to be the Fermi energy $\epsilon_F$ of Cerium, which is estimated to be 0.52eV \cite{PhysRevB.23.1266, PhysRevB.27.3390}. 

We propose this simplified model as our first attempt to incorporate the electron-phonon interaction into the study of the Cerium volume collapse. We neglect the Hubbard repulsion in the  $c$-band because it is much smaller than the Hubbard repulsion in the $4f$ band. Since Amadon and Gerossier~\cite{Amadon_2015} found that the value of the 
inter-site hopping between $4f$ electrons is less than a third of the value of the hybridization between $4f$ and conduction electrons, we also neglect the $4f$-electron inter-site hopping in our model Hamiltonian. Our goal is to construct a minimal model which displays Kondo physics and investigate the effect of introducing the electron-phonon coupling in the model. 

While this model may not provide a quantitative description of Cerium, 
it is interesting by itself as it represents an important class of many-body problems which include localized levels coupled to  correlated conduction bands   \cite{Nolting_2003,Peters_2007,Koga_2008a,Koga_2008b}. This kind of models has not been extensively studied in the literature largely due to its complexity and its associated computational challenges. Our model is particularly challenging due to the fact that 
the correlations in the conduction band due 
to electron-phonon coupling are retarded~\cite{Peng_2013}. The competition and cooperation among the Kondo effect, the Ruderman-Kittel-Kasuya-Yosida interaction, and the correlation effect in the conduction band produces a very rich phase diagram \cite{Peters_2007,Koga_2008a,Koga_2008b}. 

With the goal of understanding the volume collapse transition, in this work we  focus exclusively on the Kondo regime and investigate the effect of the electron-phonon coupling in the conduction band. In particular we calculate the total free energy to construct the phase diagram and demonstrate the first order phase transition. We find that even though our electronic model agrees with the Kondo volume collapse scenario in the absence of phonon coupling, the introduction of phonons  induces several interesting phenomena reminiscent of the Mott transition scenario. 

We solve this model using the dynamical mean field theory ~\cite{DMFT_1996}, with the continuous time quantum Monte Carlo (CT-QMC)~\cite{PhysRevB.76.035116} as our impurity solver. Since we are using a hypercubic lattice in our DMFT approximation, the bare electron DOS is Gaussian. 
Due to the fact that cerium does not display a sharp feature in the DOS near the Fermi surface, such as a flat band or a van Hove singularity,  it is unlikely that 
the choice of DOS will affect our goal of investigating the qualitative effect of electron-phonon coupling. Finally, we use the maximum entropy method \cite{jarrell1996bayesian} to extract the spectral functions from Monte Carlo simulation data. 

\section{Results}

In this section we first draw the pressure-volume ($p-{\cal V}$) phase diagram by calculating the total free energy including the contribution from the bulk modulus. We find that the first order phase transition emerges when the electron-phonon interaction is large.  For the same set of model parameters, we do not find a first order transition, even at much lower temperatures, in the absence of electron-phonon coupling.  

After constructing the phase diagram, we investigate the phase transition in more 
detail by calculating the evolution of the spectral function of the conduction band across the transition. Although we employ the periodic Anderson model, the paradigm for the KVC scenario, our results display features of a Mott transition in the conduction band due to the electron-phonon coupling. In the parameter regime where  $V$ is small, as $U_{eff}$ increases, the DOS of the $c$-electrons gradually develops a gap at the Fermi level with a width proportional to $U_{eff}$. The gap-opening in the DOS, and its proportionality to $U_{eff}$ mimics that of the Mott transition in the Hubbard model. The gap does not occur when $V$ dominates over $U_{eff}$, which compels us to argue that there is a competition between $V$ and $U_{eff}$ in our model~\cite{PhysRevLett.95.066402}. 

\subsection{Pressure-Volume Diagram and the Bulk Modulus}
Since the $\gamma\rightarrow\alpha$ transition is first order, the pressure versus volume curve develops a kink as the temperature drops below the transition point. To properly account for the static lattice contribution, we introduce a volume and temperature dependent bulk modulus term into the $p-{\cal V}$ relation~\cite{KVC_1982}. Therefore, the total pressure contains two parts, the pressure due to the electrons which we denote as $p_e$, and the pressure due to the bulk modulus term which we denote as $p_B$. 
 
\begin{figure}[h!]
\centerline{\includegraphics[scale = 0.35]{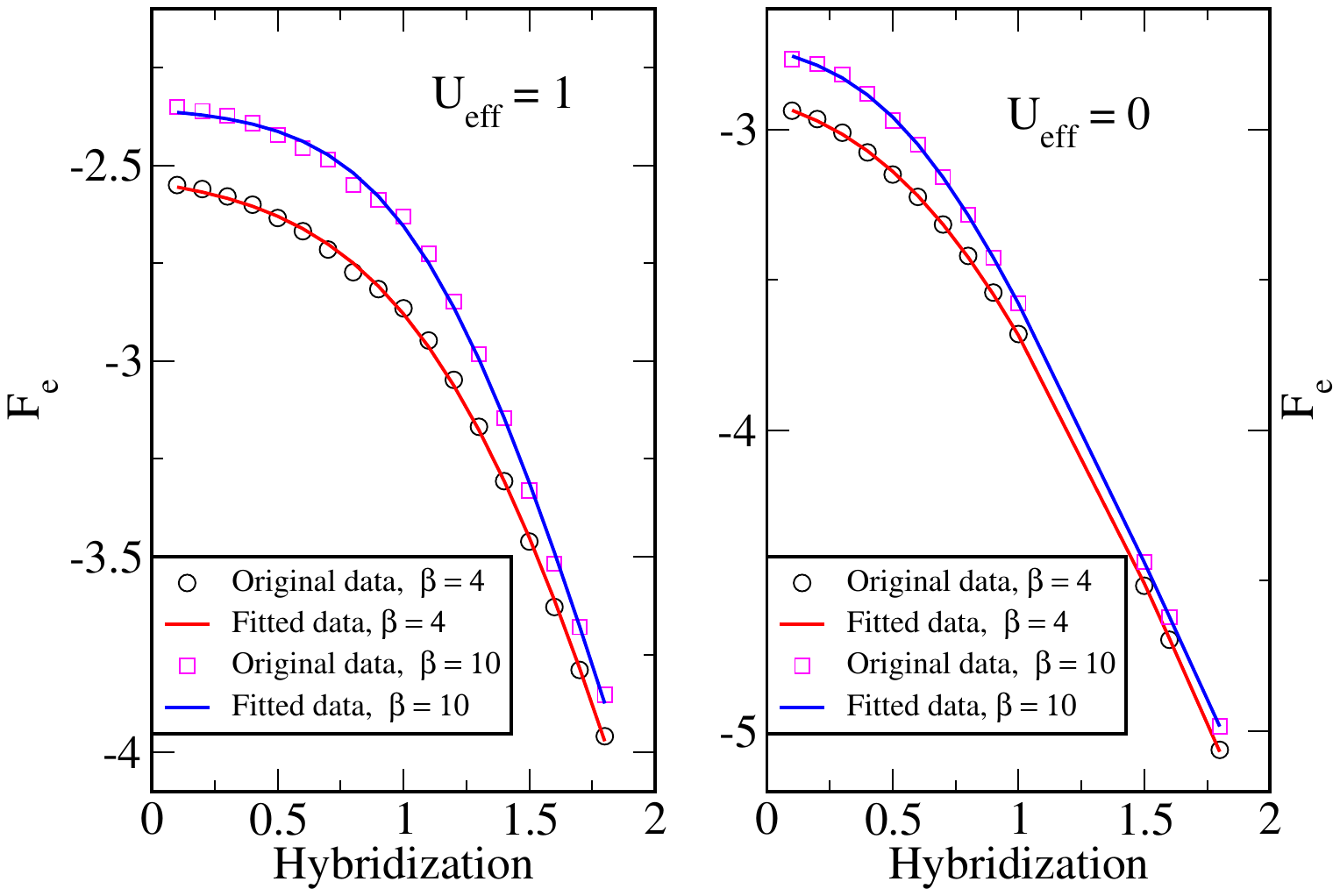}}
\caption{Left panel: Electronic free energy $F_e$ versus hybridization parameter $V$ at different temperatures for $U_{eff} = 1$. Right panel: $F_e$ versus $V$ at different temperatures for $U_{eff} = 0$. In both panels, discrete points represent the original data from numerical calculation, and the continuous curves represent our fitted results. As can be seen, most of the discrete points fall on the fitted curve, thus validating our choice of Eq. (\ref{func}). }
\label{F_V_Ueff_beta}
\end{figure}

We calculate $p_e$ from the electronic free energy by the relation $p_e = -\frac{\partial F}{\partial {\cal V}}$, and $p_B$  
by integrating  the bulk modulus $B = -{\cal V} \frac{\partial p_B}{\partial {\cal V}}$, where $\cal V$ is the volume. We calculate the electronic free energy using the formula $F_e (T = T_0, {\cal V} = {\cal V}_0, N)= \int_{0}^{N}\mu dN + F(T_0, {\cal V}_0, N=0)$. Here, we choose not to use the entropy formula employed in Ref. \cite{PhysRevB.80.140505} because the statistical error in our results
become large at high temperatures. When we plot the free energy versus hybridization $V$, we notice that the curve continuously evolves from a nearly flat plateau at small $V$ to a nearly straight line with a negative slope at large $V$, as shown in Fig. \ref{F_V_Ueff_beta}. 
For the parameter regime that we have scanned for this model, we did not find a point where the curvature of the electronic free energy versus volume 
curve changes sign, which is an indicator for the emergence of first order phase transition. 
It is this finding that motivated us to introduce the free energy due to bulk modulus to fully describe the $\gamma \rightarrow \alpha$ transition using our model, which will be discussed in detail later. 

With the observation of the curve shapes in Fig. \ref{F_V_Ueff_beta}, we conjecture that that the derivative of the free energy with respect to $V$ can be approximately fitted to the function $-k(1 + \tanh a(V-c))$, with $k, a, c$ being positive parameters. Integration of the derivative gives a function to which we can fit our free energy data:
\begin{eqnarray}
F_e(V) = -k\Big(V -c + \frac{1}{a}\log 2\cosh a (V - c)  \Big) + d. 
\label{func}
\end{eqnarray}

By fitting our numerical data to Eq.~(\ref{func}), we obtain the values of the parameters $k, a, c, d$. Since the free energy depends on the temperature, these parameters also depend on it. Now, we can obtain the volume dependence of the free energy using the empirical relationship between hybridization $V$ and volume $\cal V$,
$V = \frac{b}{{\cal V}^2}$~\cite{mcmahan1998volume}.  We calculate the electronic pressure as 
\begin{eqnarray}
p_e = -\frac{2kb}{{\cal V}^3}\Bigg( 1 + \tanh a\Big(\frac{b}{{\cal V}^2} - c\Big) \Bigg). 
\end{eqnarray}

The experimental value of $b$ can be estimated from the relation $V = \frac{b}{{\cal V}^2}$, and $J \propto \frac{V^2}{U}$\cite{PhysRev.149.491}, where $J$ is the Kondo exchange. The experimental values of $J$ range between $0.2-0.3eV$ in the $\alpha$ phase and $0.05-0.06eV$ in the $\gamma$ phase~\cite{PhysRevB.23.1266, PhysRevB.45.8934, PhysRevB.67.075108, PhysRevB.48.13981}. From the values of $J$ and the relation between $J$ and the  volume, we can estimate the value of $b$ to be between 0.89 and 1.55 in our unit system.

The second contribution to the total pressure comes from the bulk modulus. From the experimental results of Ref.~\cite{PhysRevB.7.567, murnaghan1944compressibility}, we assume that the bulk modulus depends upon the volume as
\begin{eqnarray}
B = B_{0}(T) e^{\alpha(1-{\cal V}/{\cal V}_0)}, 
\end{eqnarray}
and upon the temperature through the relation~\cite{yu2007anomalous}
\begin{eqnarray}
B_{0}(T) = B_{0}(1 + e^{-\frac{T_0}{T}}),
\end{eqnarray}
where, $B_0, T_0, {\cal V}_0$ and $\alpha$ are material-dependent parameters. 

Integration of the bulk modulus gives us the pressure $p_B$ as 
\begin{eqnarray}
p_{B}({\cal V}, T) = p_0(T) - B_0(T)\int_{1}^{{\cal V}/{\cal V}_0}dx \frac{e^{\alpha(1-x)}}{x},
\label{eq:pB}
\end{eqnarray}
where $p_0(T)$ is an arbitrary constant that may depend on the temperature. Throughout the paper, we have set $p_0 = 0$.

\begin{figure}[h!]
\centerline{\includegraphics[scale = 0.3]{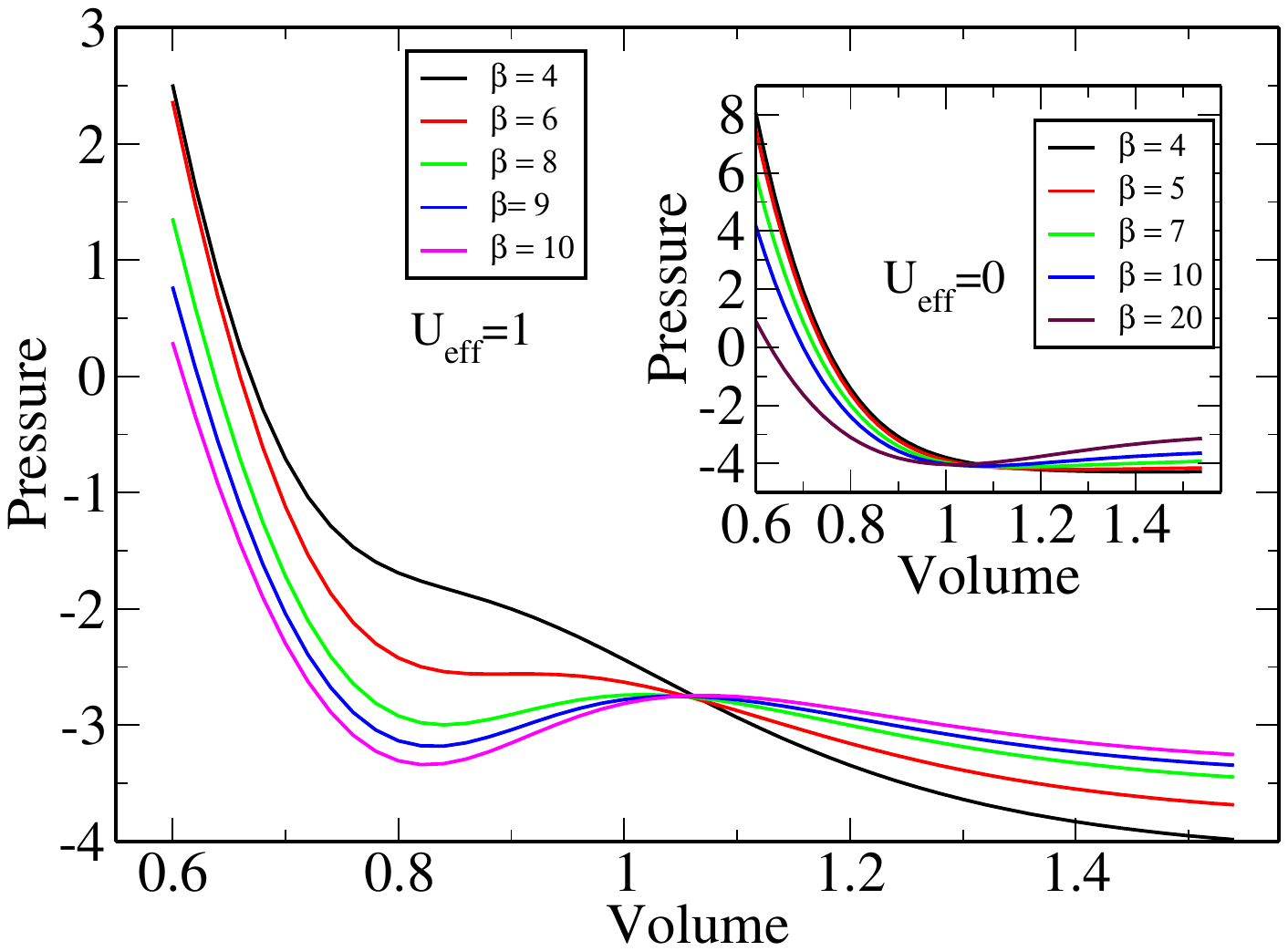}}
\caption{Main panel: The $p (\text{pressure})-{\cal V} (\text{volume})$ diagram for $U_{eff} = 1, \Omega_0 = 0.01$. As the temperature decreases (critical temperature $T_{c} = 1/6$), a kink structure develops in the $p-{\cal V}$ graph, indicating the emergence of a first order phase transition. Inset: $p-{\cal V}$ diagram for $U_{eff} = 0$. Here, with the same bulk modulus pressure, the kink structure does not show up.  }
\label{PV_Ueff}
\end{figure}

Adding the bulk modulus and the electronic pressures yields a $p-{\cal V}$ graph which exhibits a kink structure. 
Fig. \ref{PV_Ueff} shows the pressure versus volume diagrams for different values of $U_{eff}$. 
When $U_{eff} = 1$, as we lower the temperature, a kink structure begins to develop. We identify $\beta = 6$ as the critical temperature where the kink structure begins to emerge. Experimentally, the ratio between the $\gamma\rightarrow\alpha$ transition critical temperature $T_c$ and the temperature $T'$ where the volume collapse of 17\% occurs is $T_c/T' = 460/334$~\cite{moore2011watching}. Using the same ratio, we can identify the $T'$ in our model to be approximately $1/8$. From the iso-thermal $p-{\cal V}$ diagrams, we find that the volume collapse in our model at $\beta = 8$ is about 30\%, a result that is in reasonable agreement with the experiments, considering that we are using a highly simplified model. From the Maxwell construction, we can read off from the $\beta = 8$ iso-thermal line the volumes for the $\gamma$ and $\alpha$ phases, with ${\cal V}_{\alpha} = 0.78$, and ${\cal V}_{\gamma} = 1.13$. We further estimate the corresponding hybridization value for these two phases to be $V_{\alpha} = 1.93$, and $V_{\gamma} = 0.92$. 

On the other hand, when $U_{eff} = 0$ (inset on Fig. \ref{PV_Ueff}), even though we have used the same set of parameters, the kink structure that is the indicator for the emergence of a first order phase transition is absent. Note that the small upturn in the $p-{\cal V}$ diagram at large volume can also be eliminated once we consider the volume dependence for the hopping term $t$ in the conduction band.

Similar results can be obtained with  many different combinations of parameters.
In the data displayed in Fig. \ref{PV_Ueff} we set $b = 1.18, p_0=0, T_0=0.1, B_0 = 12.47$, and $\alpha = 4.225$.  The value of $b$ is within our estimated range.
A value of  $T_0=0.1$ is approximately 600 K within our units, a value comparable to the critical point temperature of the  transition. Following Ref.~\cite{KVC_1982}, we use $B_0 = 28GPa, {\cal V}_{0} = 36$\AA$^3$ as the bulk modulus and unit cell volume for Cerium in the $\gamma$ phase. 
Once we use the Fermi scale as our unit of energy and set ${\cal V}_0$ as our unit of volume, the unit of pressure becomes $\epsilon_F/{\cal V}_0 = 2.3GPa$.  This justifies our usage of the value $B_0 = 12.47$ as our bare bulk modulus.  And, finally, the experimental values of $\alpha$ range between 2 and 5 for most bulk pure metals~\cite{PhysRevB.7.567}.

In summary, for a large range of parameters, if the  electron-phonon coupling ($U_{eff} = 1$) is finite, there is a clear first order
transition in the $p- {\cal V}$ diagram
with a critical temperature around $1/6$. However, for the same set of parameters, when the electron-phonon interaction is absent ($U_{eff}=0$), 
the transition is not seen for temperatures down to $1/20$. 
The different behavior of the $p-{\cal V}$ diagram for $U_{eff} = 0$ and $U_{eff} = 1$ implies that the electron-phonon interaction enhances the $\gamma\rightarrow\alpha$ volume collapse transition.

\subsection{Spectral Functions of the Mott Metal-Insulator Transition}
\begin{figure}[h!]
\centerline{\includegraphics[scale = 0.3]{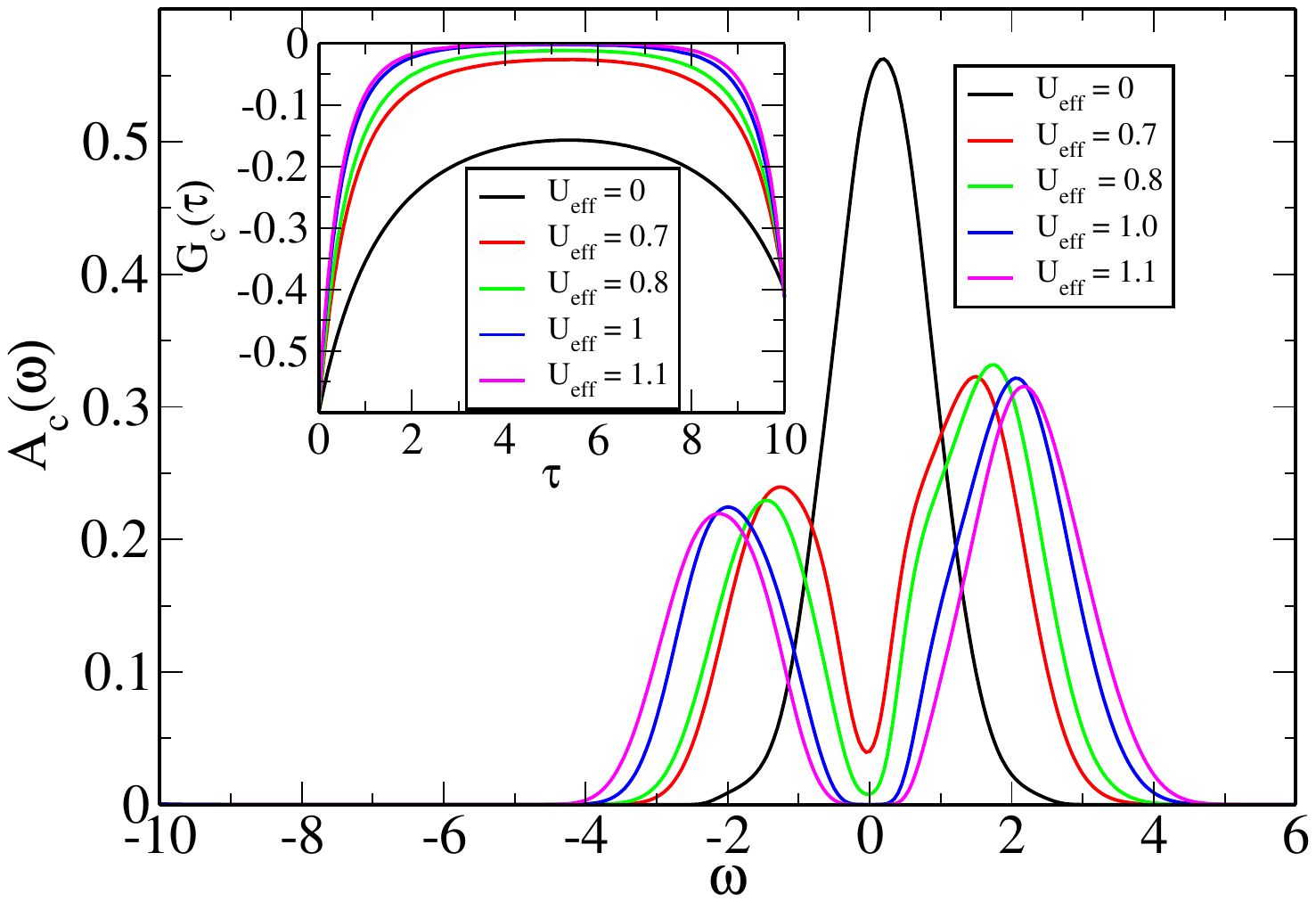}}
\caption{Main panel: The conduction electron spectral functions for $V = 0.1, \beta = 10, \Omega_0 = 0.01$. As $U_{eff}$ increases from 0.0 to 1.1, the gap in the DOS becomes increasingly wider. Inset: The evolution of the local conduction electron Green's function $G_c(\tau)$ with $U_{eff}$. The increasingly rapid decay of $G_c(\tau=\beta/2)$ also indicates the existence of an energy gap at $\omega = 0$ as $U_{eff}$ grows. }
\label{smallV}
\end{figure}

The phonon-enhanced first order phase transition can be interpreted as a Mott metal-insulator transition. We can understand this by studying the evolution of the spectral functions with respect to the variation of the relative strengths of $V$ and $U_{eff}$. When the hybridization is small, the electron-phonon interaction can significantly modify the density of states (DOS) of the conduction electrons. For $V = 0.1$, as $U_{eff}$ increases from 0 to 1.1, the DOS changes from a nearly Gaussian to a DOS  gapped at the Fermi energy, as shown in Fig. \ref{smallV}. The gap in the conduction electron spectral function is not an artifact of the maximum entropy method since 
we can also see the effect of $U_{eff}$ on the gap at the Fermi energy by observing the behavior of the local $c$-electron Green's function $G_c(\tau)$ (inset of Fig.~\ref{smallV}),
which is directly measured in the Monte Carlo simulation.   
When there is no gap, the value of $G_c(\tau = \beta/2)$ is finite. However, when there is a gap at $\omega = 0$, the value of  $G_c(\tau = \beta/2)$ decays to zero exponentially. Moreover, the wider the gap, the more rapid the decay. When we plot the $G_c(\tau)$ for different values of $U_{eff}$, we see clearly that with increasing $U_{eff}$, $G_c(\tau)$ decreases increasingly rapidly when $\tau$ approaches $\beta/2$. 

The magnitude of the phonon frequency $\Omega_0$ and the $c$-electron filling have profound influences on the nature of the Mott transition due to $U_{eff}$. When $\Omega_0$ is small, which is the case we are studying, the transition seems to be continuous or at most weakly discontinuous~\cite{meyer2002gap}. The reason is that when $\Omega_0 = 0$, we can integrate out the Holstein phonons in the conduction band to obtain a generalized Falicov-Kimball model (A Falicov-Kimball model  with $c-f$ hybridization) in which the conduction electrons always exhibit a non-Fermi liquid behavior for any non-trivial filling number~\cite{PhysRevB.46.1261, freericks2003exact}. The  Mott transition from a non-Fermi liquid metal to an insulator is continuous due to the absence of the quasi-particle peak at the Fermi energy in the conduction electron DOS ~\cite{muller1989hubbard}. Notice here we 
are considering only the transition in the electron-phonon system with model Hamiltonian given by 
Eq.~\ref{eq:Hamiltonian}. By including the static lattice contribution this 
transition becomes first order as we discussed earlier. 

 A small phonon frequency is chosen in this study because of the low Debye temperature observed in experiments  ~\cite{phonon_2009, phonon_2011, entropy_2006,Jeong_2004}. 
The low phonon frequency greatly reduces the pairing instability. With this phonon frequency and a small enough hybridization between the conduction and the localized $4f$ electrons, we find that the charge density wave (CDW) instability always dominates, which is consistent with previous results using the Holstein model and DMFT \cite{Mark_1993}. 

\begin{figure}[h!]
\centerline{\includegraphics[scale = 0.3]{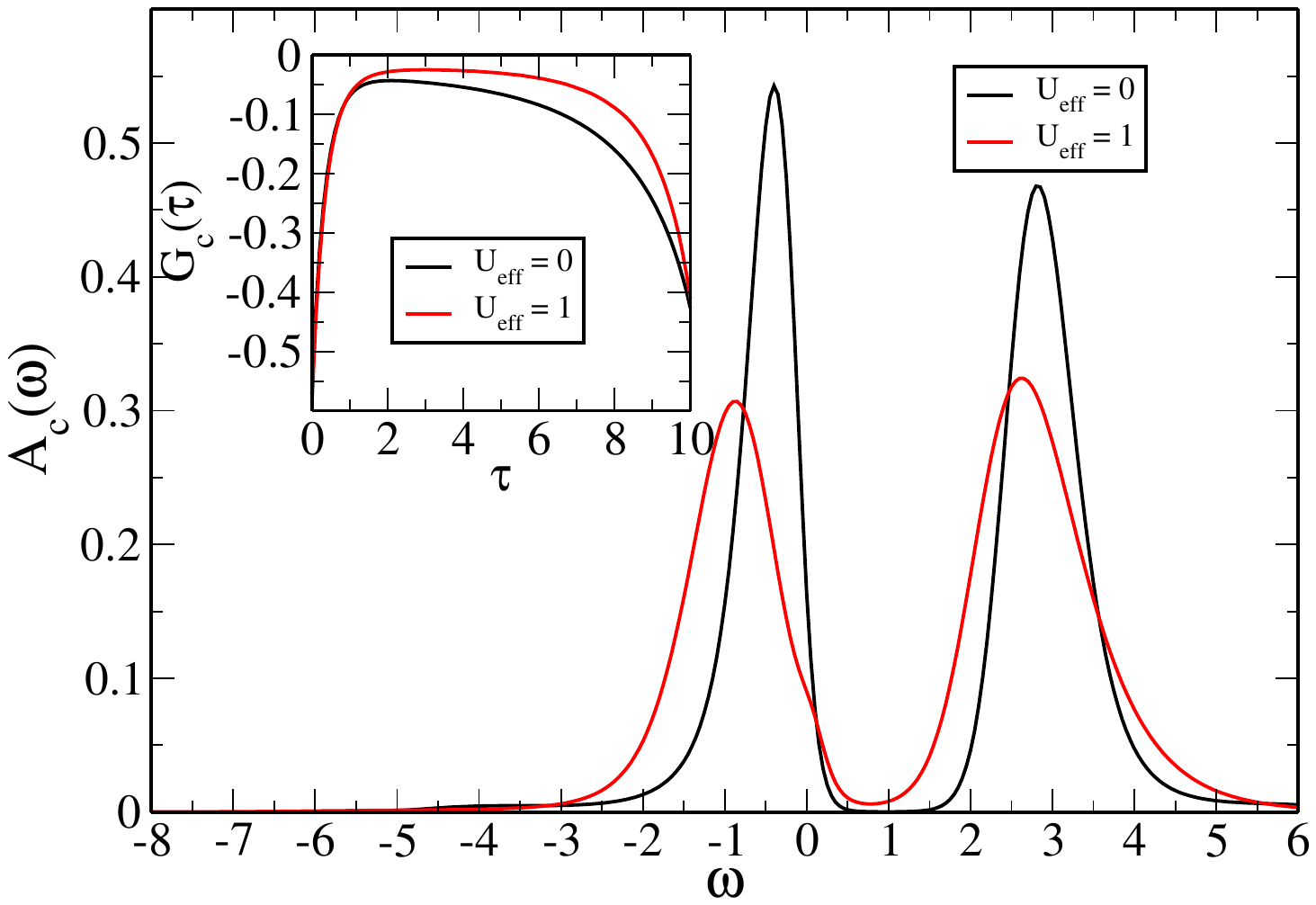}}
\caption{Main panel: The conduction electrons DOS for $U_{eff} = 0$ and $U_{eff} = 1$ when $V = 1.8,  \beta = 10$, $\Omega_0 =0.01$. Here, the introduction of the electron-phonon interaction has no significant influence on the DOS, which at the Fermi energy is always finite whether there is electron-phonon interaction or not. Inset: The local $G_c(\tau)$ for small and large electron-phonon interaction strength.}
\label{largeV}
\end{figure}

The opening of the Mott gap at the Fermi level is present only when the hybridization between conduction band and localized electrons is weak compared with the electron-phonon coupling.  Fig.~\ref{largeV} shows that when the hybridization is strong, the opening of the Mott gap is prohibited. In this parameter regime, the introduction of the electron-phonon interaction has little effect on the behavior of the conduction electron DOS. Since the filling number of the $c$-electrons is set to 0.8, the hybridization cannot induce a gap at the Fermi energy~\cite{pruschke2000low}, and thus the DOS is always finite irrespective whether there is electron-phonon interaction or not. Consequently, the $c$-electrons are always metallic in the large $V$ regime. 

\begin{figure}[h!]
\centerline{\includegraphics[scale = 0.3]{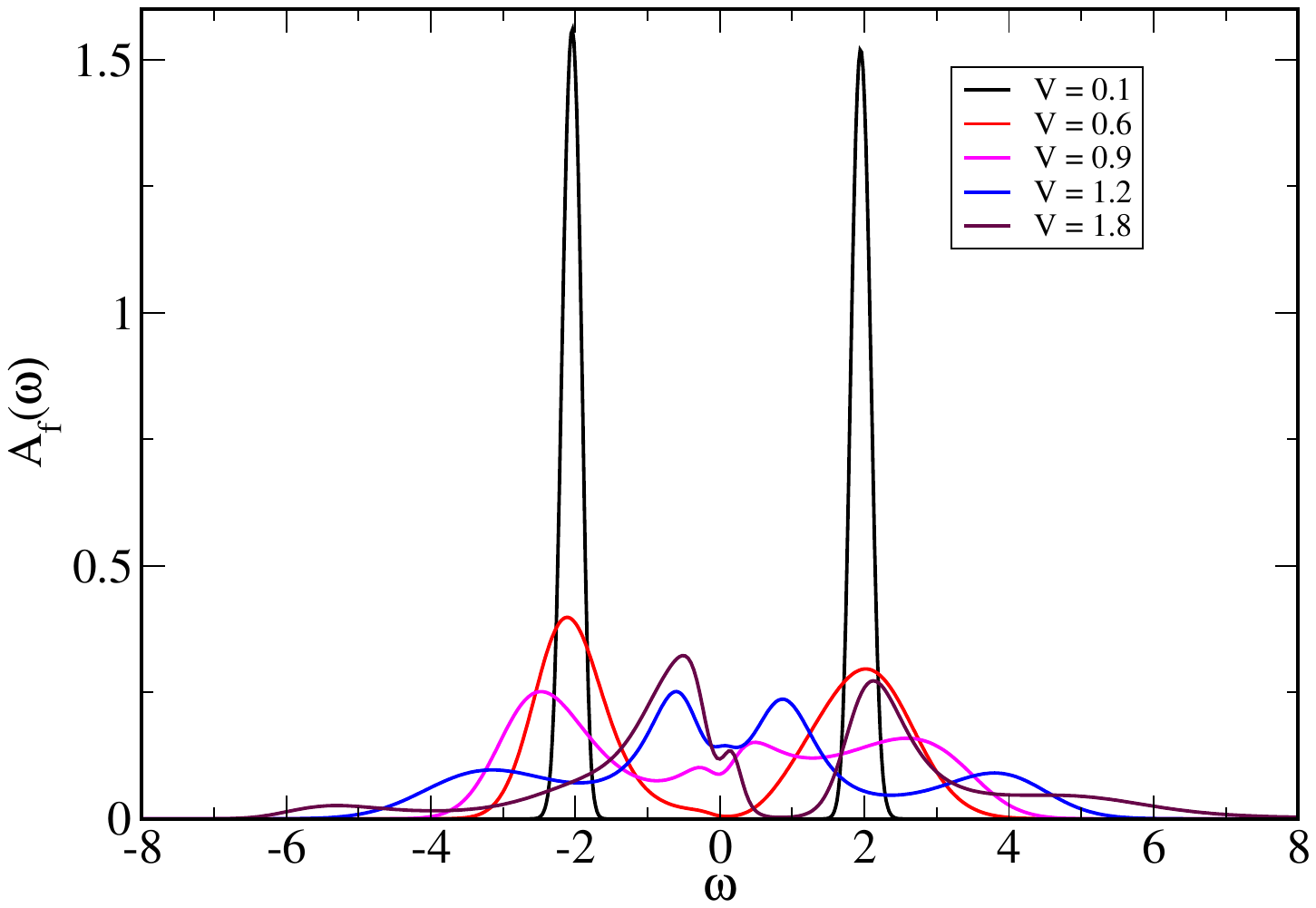}}
\caption{The $f$-electron spectral functions for $\beta = 10, \Omega_0 =0.01, U_{eff} = 1$,  for various hybridization values ranging from $V = 0.1$ ($\gamma$ phase) to $V = 1.8$ ($\alpha$ phase). When $V$ is small, the electron-phonon interaction dominates over the Kondo effect, and the $f$ electron spectral function shows no signature of Kondo resonance. When $V$ gets increasingly larger, the Kondo effect kicks in, which can be seen from the Kondo resonance that is already developing at $\beta = 10$. }
\label{ff_spectra}
\end{figure}

The absence of the Mott gap in the large $V$ regime signals that the electron-phonon interaction effect is suppressed by the hybridization. Since the electron charge susceptibility is positively correlated with $U_{eff}$, the suppression of the electron-phonon coupling effect is also reflected in the decrease of the charge susceptibility as $V$ increases for fixed $\beta$ (not shown).  As $V$ increases from 0.1 to 1.8, the localized $f$ electron moments that are present at small $V$ get screened by the conduction electrons when $V$ is large \cite{Peng_2013}. The screening of the localized $f$ electron moments in the large $V$ regime is a signature of the Kondo effect, which we can also observe by studying the $f$ electron spectral functions. Fig. \ref{ff_spectra} shows the $f$ electron spectral functions for small $V$ ($\gamma$ phase) and large $V$ ($\alpha$ phase) at $\beta = 10$. As we can see in the figure, when $V$ is small, Kondo temperature $T_{K}$ is much smaller than $\beta = 10$, and the Kondo resonance is nowhere to be found in the $f$ electron spectral curves. However, as $V$ increases, Kondo temperature $T_{K}$ also gets larger, which we can see from the Kondo resonances that emerge for large $V$. The absence of the Kondo resonance for small $V$ and its emergence at large $V$ is consistent with the experimental observation that $T_{K}$ at $\gamma$ phase is much smaller than $T_{K}$ at $\alpha$ phase. We also notice that the $f$ electron spectral curve gets kinky at $\omega = 0$ when $V = 1.8$. The strength of this kink is proportional to the effective electron-phonon interaction strength $U_{eff}$, and thus we conjecture that the competition between $c-f$ hybridization $V$ and $U_{eff}$ could lead to some exotic behavior in the $f$ electron spectral functions. The nature of this kink is still under study. 

\begin{figure}[h!]
\centerline{\includegraphics[scale = 0.3]{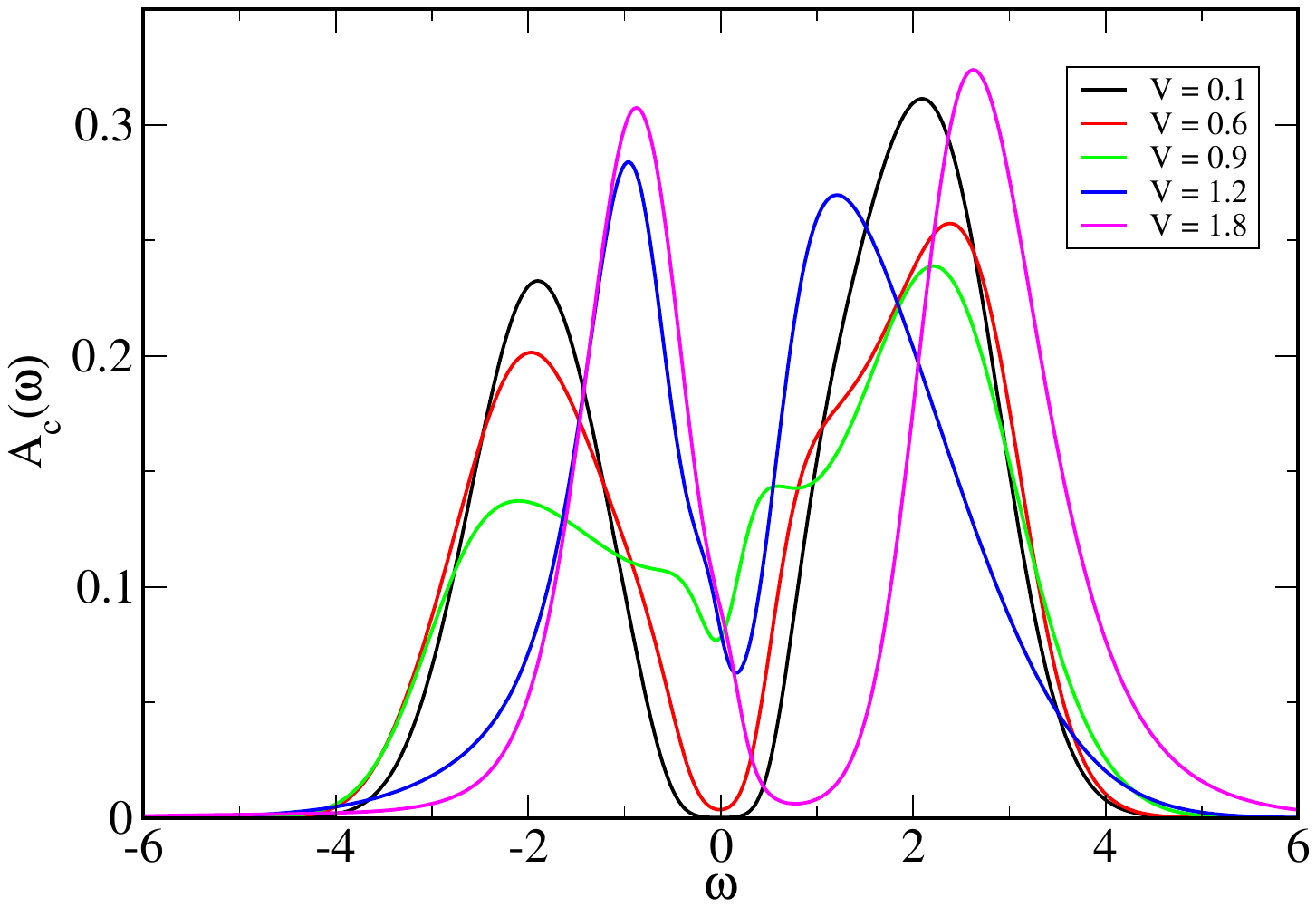}}
\caption{The $c$-electron spectral functions for $\beta = 10, \Omega_0 =0.01, U_{eff} = 1$,   at different values of $V$. When $V$ is small (below 0.6), the electron-phonon interaction dominates over the Kondo screening, and the conduction electrons form a Mott insulator. When $V$ gets large, the Kondo effect dominates over the electron-phonon interaction effect, and the conduction electrons become metallic. }
\label{dos_V}
\end{figure}

At the same time, when we scan $V$ from $V = 0.1$ to $V = 1.8$, the conduction electrons make a transition from insulator to metal. The $c$-electron
spectral functions for $\beta = 10$ and $U_{eff} = 1$ with varying values of $V$ are shown in Fig. \ref{dos_V}, where a gap at the Fermi energy is clearly visible for $V < 0.6$.  When $V > 0.6$, the Mott gap evolves into a depression which disappears completely for $V > 1.2$. Therefore the Mott metal-insulator transition is present only when the electron-phonon interaction is strong enough.

\section{Conclusion}
In brief, using the periodic Anderson model with phonon coupled to the conduction band, and by introducing a volume and temperature dependent bulk modulus contribution to the total pressure, we find a first order phase transition where the Cerium volume collapses. This transition is enhanced by the presence of the electron-phonon interaction; with other parameters being fixed, we do not find the first order transition in the absence of the electron-phonon interaction. Our findings support recent experimental results showing that phonons play an important role in the volume collapse transition in Cerium.  

Moreover, we find that our model, although originally conceived with the Kondo volume collapse scenario in mind, exhibits interesting features of the metal to insulator transition, e.g., a gap proportional to the effective electron-phonon interaction, $U_{eff}$, 
opens at the Fermi energy at low temperature. 

An obvious improvement over the present work is to include other contributions to the model. The spin orbit coupling and the hybridization in the f-band can be considered in the DMFT calculation, however a more realistic electron-phonon coupling is currently beyond the formalism of DMFT since this requires a numerical method which can handle long range coupling properly. 

\textbf{Acknowledgements}. This material is based upon work supported by the National Science Foundation under the NSF EPSCoR Cooperative Agreement No. EPS-1003897 with additional support from the Louisiana Board of Regents. 
MJ was also supported by the NSF  Materials Theory grant DMR1728457.
Computer support is provided by the Louisiana Optical Network Initiative, and by HPC@LSU computing. We would like to thank F.F. Assad for the continuous time quantum Monte Carlo programs that he shared with us. M. J. designed and implemented the maximum entropy method algorithm for extracting real-frequency data from imaginary frequency results.

\bibliography{ref}
\bibliographystyle{apsrev}

\end{document}